# Fast Whole-Brain MR Multi-Parametric Mapping with Scan-Specific Self-Supervised Networks

Amir Heydari[a], Abbas Ahmadi[a,*], Tae Hyung Kim[b], Berkin Bilgic[c,d]

[a] *Department of Industrial Engineering and Management Systems, Amirkabir University of Technology, Tehran, Iran.*

[b] *Department of Computer Engineering, Hongik University, Seoul, Korea*

[c] *Athinoula A. Martinos Center for Biomedical Imaging, Massachusetts General Hospital, Charlestown, MA, United States*

[d] *Radiology, Harvard Medical School, Boston, MA, United States*

\* Email: abbas.ahmadi@aut.ac.ir

**Abstract**

Quantification of tissue parameters using MRI is emerging as a powerful tool in clinical diagnosis and research studies. The need for multiple long scans with different acquisition parameters prohibits quantitative MRI from reaching widespread adoption in routine clinical and research exams. Accelerated parameter mapping techniques leverage parallel imaging, signal modelling and deep learning to offer more practical quantitative MRI acquisitions. However, the achievable acceleration and the quality of maps are often limited. Joint MAPLE is a recent state-of-the-art multi-parametric and scan-specific parameter mapping technique with promising performance at high acceleration rates. It synergistically combines parallel imaging, model-based and machine learning approaches for joint mapping of $T_1$, $T_2^*$, proton density and the field inhomogeneity. However, Joint MAPLE suffers from prohibitively long reconstruction time to estimate the maps from a multi-echo, multi-flip angle (MEMFA) dataset at high resolution in a scan-specific manner. In this work, we propose a faster version of Joint MAPLE which retains the mapping performance of the original version. Coil compression, random slice selection, parameter-specific learning rates and transfer learning are synergistically combined in the proposed framework. It speeds-up the reconstruction time up to 700 times than the original version and processes a whole-brain MEMFA dataset in 21 minutes on average, which originally requires ~260 hours for Joint MAPLE. The mapping performance of the proposed framework is ~2-fold better than the standard and the state-of-the-art evaluated reconstruction techniques on average in terms of the root mean squared error.

**Keywords:** quantitative MRI, parameter mapping, scan-specific deep learning, self-supervised networks

## 1. Introduction

Measurement of biophysical properties, such as spin-lattice relaxation time ($T_1$) and spin-spin relaxation time ($T_2$) as indicators of the biological state of the tissue and the microstructural processes during aging and diseases like multiple-sclerosis, brain tumors, epilepsy, stroke and iron overload is considered to be critical for diagnosis and brain research (Heydari et al., 2024; Huizinga et al., 2016; Lu et al., 2024; Nachmani et al., 2019). Moreover, quantitative MR imaging (qMRI) promises to reduce the sensitivity to hardware imperfections and improve the reproducibility and comparability of the results (Byanju et al., 2022; Kofler et al., 2024; Weiskopf et al., 2013). It allows many of the site-specific effects to be factored out or mitigated, increasing the comparability of measures across multiple sites (Balbastre et al., 2021; Bauer et al., 2010; Weiskopf et al., 2013).

qMRI has not yet been established for widespread application in clinical diagnostics (Byanju et al., 2022; Jun et al., 2021a; Kofler et al., 2024; Petzschner et al., 2011). To quantify tissue properties, distinct MRI experiments are typically performed to probe each individual property by acquiring multiple contrasts in which one or a few experimental imaging parameters like repetition time (TR), echo time (TE) or flip angle (FA) are varied, which may result in increased scan times (Feng et al., 2022; Nachmani et al., 2019; Planchuelo-Gómez et al., 2024). Long scan times make the acquisition more susceptible to the patient motion and other physiologic noise sources. Post-processing steps are often required to fit the acquired image series to a signal model/dictionary so that the corresponding MR parameter can be estimated (Feng et al., 2022; Liu and Kijowski, 2017). Depending on the number of volumes in the series to be processed, the complexity/size of the signal model/dictionary and the number of MR parameters to be mapped, the post-processing step can potentially take a large proportion of the reconstruction time.

To meet the clinical and research requirements, there has been remarkable advances in qMRI in terms of scan times, reconstruction techniques and robustness in recent years. This is in conjunction with dramatic improvement in speed of MR imaging with faster pulse sequences, more efficient sampling trajectories, better gradient systems and coil arrays (Feng et al., 2022). New acquisition techniques for parameter quantification are capable of multi-parameter mapping of whole brain in a few minutes at various resolutions (Deoni et al., 2005; Gulani et al., 2004; Preibisch and Deichmann, 2009; Schmitt et al., 2004; Warntjes et al., 2007). Accelerated imaging techniques for reconstruction of under-sampled MR data particularly in parallel imaging (Breuer et al., 2005; Griswold et al., 2002; Kellman et al., 2001; Pruessmann et al., 1999; Sodickson and Manning, 1997; Sodickson and McKenzie, 2001) and spatiotemporal acceleration techniques (Huang et al., 2005; Tsao et al., 2003; Tsao and Kozerke, 2012; Xu et al., 2007) have been demonstrated for rapid qMRI with improved reconstruction performance (Doneva et al., 2010; Feng et al., 2011; Huang et al., 2012; Zhang et al., 2015a). The use of parallel imaging techniques in qMRI traditionally follows separate steps for reconstruction of under-sampled data and parameter fitting. On the contrary, model-based reconstruction techniques are capable of direct estimation of MR parameters from acquired data (Block et al., 2009; Fessler, 2010; Sumpf et al., 2011; Wang et al., 2019, 2018).

Parallel imaging can fit accelerated, under-sampled acquisitions by exploiting the complementary information provided by different sensitivity profiles of coil arrays (Hamilton et al., 2017). Furthermore, multi-contrast nature of data in qMRI enables the use of joint reconstruction techniques (Byanju et al., 2022; Haldar, 2013; Kim et al., 2019; Polak et al., 2020; Yaman et al., 2021; Zhang et al., 2015b) which are exploiting the relationships across contrasts and thereby permitting higher acceleration rates. However, the achievable acceleration rate may be limited (Sardá-Espinosa, 2019; Vasanawala et al., 2010) and processing more information in multi-channel, multi-contrast dataset comes with increased computational costs.

Model-based MR parameter mapping techniques synergistically combine the image reconstruction and parameter fitting steps which allows them to directly estimate MR parameters from under-sampled data, leading to significant imaging efficiency. Model-based techniques exploit the relationship between acquired data and a signal model containing the parameters of interest and cast it in the form of an inverse problem which can be augmented with different regularization strategies. These techniques are more interpretable, reliable and generalizable (Feng et al., 2022; Lu et al., 2024; Maier et al., 2019).

Machine learning tools have created additional gains in qMRI by incorporating deep learning architectures like CNN (Chaudhari et al., 2018; Li et al., 2020), FCN (Cohen et al., 2018; Liu et al., 2020), U-Net (Fang et al., 2019; Liu et al., 2019), ResNet (Cai et al., 2018; Zhang et al., 2019) , GAN (Liu et al., 2020) to improve the existing parallel imaging reconstruction and model-based techniques (Jun et al., 2021b; Monga et al., 2021; Sabidussi et al., 2021) or directly map the parameters from acquired data in an end-to-end manner (Cai et al., 2018; Li et al., 2020). They improve the accuracy and efficiency of qMRI but may lack interpretability and generalizability. The need for large datasets in the training process may hamper the applicability of some deep learning methods. Scan-specific and self-supervised methods (Heydari et al., 2024; Jun et al., 2024; Kim et al., 2022) are developed to mitigate the need for large datasets and improve the applicability of deep learning in qMRI.

Early attempts in qMRI aim to estimate a single parameter at a time using sequences with carefully chosen parameters (Gupta, 1977). The need for translating qMRI to clinical and research applications has led to the development of tailored sequences for multi-parameter mapping (Balbastre et al., 2021; Jun et al., 2024; Seiler et al., 2021; Weiskopf et al., 2013). MR fingerprinting (Ma et al., 2013) allows multi-parameter mapping with a single acquisition without the need to reconstruct dynamic clean images (Hamilton and Seiberlich, 2019; Jaubert et al., 2020; Lima da Cruz et al., 2019). However, the large size of dictionaries in cases where there is an increased number of MR parameters to be estimated is one of the challenges of this approach.

There have been recent attempts to combine different approaches of parallel imaging reconstruction, model-based, deep learning and multi-parametric MR parameter mapping which offer additional gains in terms of efficiency and accuracy. Kim et al. introduced the MAPLE framework (MR-accelerated parameter mapping with cyclic loss and unsupervised scan-specific networks) which synergistically combines parallel imaging reconstruction and model-based MR parameter mapping (Kim et al., 2022). MAPLE can incorporate modern parallel imaging reconstruction methods and exploit relaxation models that can be expressed analytically. Recently, Heydari et al. proposed a multi-parametric extension of MAPLE called Joint MAPLE with significant improvement in fidelity of MR parameter mapping at acceleration rates as high as 16 and 25-fold (Heydari et al., 2024). Joint MAPLE uses scan-specific, self-supervised networks in its parallel imaging reconstruction block without the need for large datasets and is capable of jointly reconstructing multi-contrast data. It utilises MEMFA dataset in its training and optimization process, enabling higher acceleration rates while estimating high-fidelity maps. However, processing a high-resolution, volumetric MEMFA data and using a complex signal model require long computation times.

In this work, we propose a fast version of Joint MAPLE which estimates $T_1$, $T_2^*$, proton density and frequency maps jointly for the whole-brain and is up to 700 times faster than the original Joint MAPLE on average. We call it FTL-Joint MAPLE (Fast Transfer Learning-Joint

MAPLE) which uses coil compression, a random slice selection strategy, parameter-specific learning rates and transfer learning synergistically.

The proposed FTL-Joint MAPLE selects a random subset of slices in each epoch during the training/fine-tuning phase instead of processing all slices within the volume. The random slice selection avoids the need for exhaustive training of the network for all slices in each iteration while retaining the generalizability of the outputs to all slices. The size of a multi-channel dataset can be reduced using coil compression. These techniques allow for significant reduction in the amount of data that need to be processed, so that parallel imaging techniques can become more computationally efficient.

Adoption of different learning rates in machine learning has been introduced as an effective way to increase the accuracy and training speed of deep learning applications. It generally addresses the problems like vanishing gradients in complex deep and multi-layer structures, low-curvature saddle points (Singh et al., 2015) and multimodal fusion in multi-modality deep learning techniques (Yao and Mihalcea, 2022). The optimization process for MR parameter mapping in Joint MAPLE suffers similar problems leading to a slow convergence with different convergence rates for different MR parameters. It optimizes $T_1$, $T_2^*$, proton density and frequency maps with different dynamic ranges in a complex signal model. The parameter values are updated voxel-by-voxel by back-propagating different loss terms where a large amount of data is being evaluated in each loss term.

Employment of transfer learning in medical image analysis aims to address the generalization problem of deep learning methods as they often need large numbers of samples to train the network and to achieve robust generalization (Lv et al., 2021; Valverde et al., 2021). Since large training datasets may be difficult to obtain in clinical applications, many deep learning techniques use publicly available datasets to pretrain the network and then seek to generalize its application to different tasks using fine tuning (Arshad et al., 2021; Dar et al., 2020; Han et al., 2018; Knoll et al., 2019). The proposed framework employs a simple transfer learning approach to accelerate the convergence of the reconstruction by initialization of the network parameters.

Our experiments show that while reaching high speed in reconstruction time, FTL-Joint MAPLE retains the accuracy of the original Joint MAPLE. It also significantly outperforms other standard and modern parallel imaging reconstruction methods in terms of both mapping time and the accuracy of the results.

Data/Code is available on*: https://github.com/AmirHeydariGit/fast_joint_maple*

## 2. Methods

### 2.1. *Framework*

The proposed FTL-Joint MAPLE extends Joint MAPLE by adapting it for whole-brain reconstruction, incorporation of coil compression, using parameter-specific learning rates in the optimization process and exploiting a transfer learning approach.

Joint MAPLE (Fig. 1) estimates $T_1$, $T_2^*$, proton density and frequency maps jointly in a plug and play manner between a parallel imaging reconstruction block (recon block) and multi-parameter model block (model block). The joint ZS-SSL (Yaman et al., 2021) (zero shot self-supervised learning) $R_\theta$, embedded in the recon block, jointly reconstructs under-sampled MEMFA k-space data $d$.

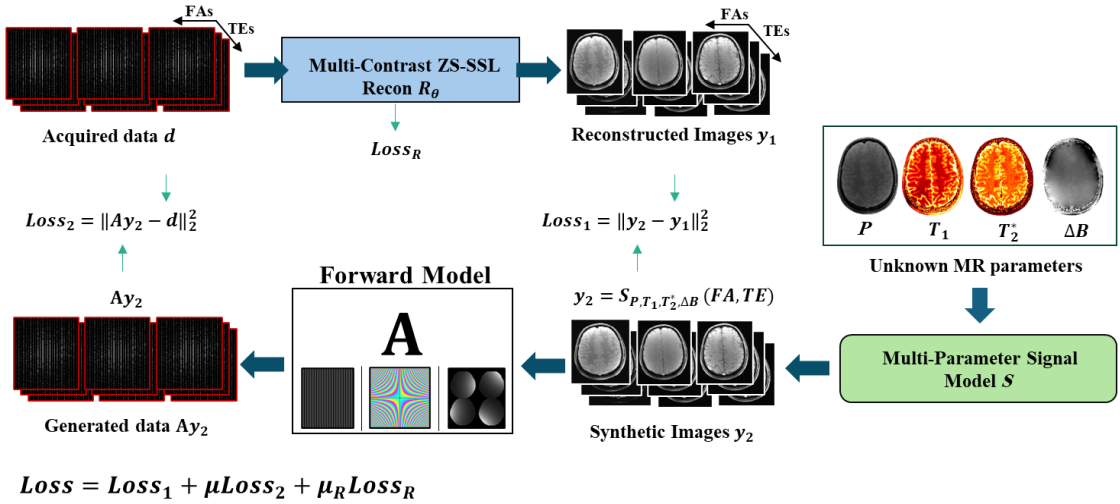

**Fig. 1. Joint MAPLE framework.** Joint ZS-SSL reconstructs under-sampled multi-echo, multi-flip angle k-space data (recon block) and multi-parameter signal model generates synthetic images using the unknown MR maps (model block). Matching reconstructed and synthesized contrasts form *Loss₁* and matching k-space related to the synthetic images (using the forward model **A**) and acquired data forms *Loss₂*. The unknown MR parameters are estimated in a plug-and-play manner between recon and model block with minimization of Loss function, which is a scaled summation of *Loss₁*, *Loss₂* and the training loss term of the joint ZS-SSL network *LossR*.

$$\boldsymbol{y_1} = R_\theta(\boldsymbol{d}) \tag{1}$$

$y_1$ denotes the reconstructed multi-contrast images and $\theta$ is the set of network parameters. The model block uses a multi-parametric signal model $S$, containing the unknown $\mathbf{T_1}$ and $\mathbf{T_2^*} \in \mathbb{R}$, $\mathbf{P} \in \mathbb{C}$ (proton density) and magnetic field inhomogeneity $\Delta\mathbf{B} \in \mathbb{R}$ maps, to generate synthesized multi-contrast images $y_2$ which is a function of the sequence parameters of TEs and the FAs.

$$y_2 = S_{P,T_1,T_2^*,\Delta B}(\text{FA}, \text{TE}) = \tag{2}$$

$$\left[\mathbf{P}\exp\left(-\frac{\text{TE}}{\mathbf{T_2^*}}\right)\exp(i2\pi\Delta\mathbf{B}\cdot\text{TE})\right]\sin(\text{FA})\frac{1-\exp\left(-\frac{\text{TR}}{\mathbf{T_1}}\right)}{1-\cos(\text{FA})\exp\left(-\frac{\text{TR}}{\mathbf{T_1}}\right)}$$

In comparison to the Joint MAPLE, we ignored transmit inhomogeneity $B_1^+$ to simplify the signal model. For parallel imaging and model-based techniques that use individual signal models for separate estimation of $T_1$ and $T_2^*$, Eq. (3) and Eq. (4) can be used:

$$S_{P,T_1}(\text{FA}) = \mathbf{P}\sin(\text{FA})\frac{1-\exp\left(-\frac{\text{TR}}{\mathbf{T_1}}\right)}{1-\cos(\text{FA})\exp\left(-\frac{\text{TR}}{\mathbf{T_1}}\right)}\exp(-i\emptyset_{est}) \tag{3}$$

$$S_{P,T_2^*,\Delta B}(\text{TE}) = \mathbf{P}\exp\left(-\frac{\text{TE}}{\mathbf{T_2^*}}\right)\exp(i2\pi\Delta\mathbf{B}\cdot\text{TE}) \tag{4}$$

$\emptyset_{est}$ is an estimated phase to compensate for potential model mismatch across different flip angles. Using Eq. (3) for $T_1$ mapping does not require different TEs. Contrasts from each TE with different FAs can allow for a separate estimation of $T_1$. Likewise, $T_2^*$ mapping with the signal model in Eq. (4) only needs different contrasts of each flip angle with different TEs to generate separate $T_2^*$ maps for each FA. In other words, when using individual signal models of Eqs. (3) and (4), a MEMFA dataset with $N_{TE}$ echo times and $N_{FA}$ flip angles can estimate $N_{TE}$ $T_1$ maps and $N_{FA}$ $T_2^*$ maps.

The output of the recon block in Joint MAPLE is required to be consistent with the synthesized multi-contrast images of the signal model which forms the first loss function ($Loss_1$).

$$Loss_1 = \|y_2 - y_1\|_2^2 \tag{5}$$

Using a forward model **A** consisting of coil sensitivities, Fourier transform and under-sampling masks, the k-space related to the synthetic images can be matched with the acquired MEMFA data and enforces the second data consistency term $Loss_2$ in k-space.

$$Loss_2 = \|\mathbf{A}y_2 - d\|_2^2 \tag{6}$$

The joint ZS-SSL loss function ($Loss_R$) is also included in the total loss term to involve the training of the network parameters for multi-contrast image reconstruction in MR parameter mapping (Heydari et al., 2024; Yaman et al., 2021),

$$Loss_R = \frac{1}{KN}\sum_{t=1}^{N}\sum_{k=1}^{K} L\left(d_{\Lambda_k^t}^t, \mathbf{A}_{\Lambda_k^t}\left(g\left(d_{\Gamma_k^t}^t, \mathbf{A}_{\Gamma_k^t}; \theta\right)\right)\right) \tag{7}$$

where $L$ is an $l_1$-$l_2$ loss defined between the output of the network and a reference acquired k-space data. $t$ is the index for N contrasts (N = $N_{TE} \times N_{FA}$) and $k$ is the index for different subsets/combinations of all K partitions in the multi-mask setting of ZS-SSL. In Eq. (7) $g$ denotes the unrolled network operating with $\mathbf{A}_{\Gamma_k^t}$ (the forward model consisting of $k$th training mask specific for contrast t ($\Gamma_k^t$)) and $d_{\Gamma_k^t}^t$ (the acquired k-space of contrast $t$ under-sampled by $\Gamma_k^t$) to output reconstructed contrasts. The forward model $\mathbf{A}_{\Lambda_k^t}$ including the loss mask ($\Lambda$) operates on the output of the unrolled network to match the acquired k-space $d_{\Gamma_k^t}^t$ under-sampled with the loss mask. Please refer to Heydari et. al for further details on $Loss_R$. The scaled summation of $Loss_1$, $Loss_2$ and $Loss_R$ creates the final loss function $Loss$ which jointly optimizes $T_1$, $T_2^*$, P, $\Delta$B as MR parameters and $\theta$ as the network parameters.

$$\mathbf{P}, \mathbf{T}_1, \mathbf{T}_2^*, \Delta\mathbf{B}, \theta = \underset{\mathbf{P}, \mathbf{T}_1, \mathbf{T}_2^*, \Delta\mathbf{B}, \theta}{\operatorname{argmin}} Loss = Loss_1 + \mu Loss_2 + \mu_R Loss_R \tag{8}$$

The contribution of each loss term is adjusted with regularization weights of $\mu$ and $\mu_R$. The results for evaluated subjects show that a value of 100 for $\mu$ and 0.001 for $\mu_R$ can yield near-optimal outputs. However, based on a heuristic approach, a set of values for each regularization weight can be evaluated to minimize NRMSE and obtain better reconstruction and mapping performance. In the supplementary Fig. S8 the effect of applying constant values of $\mu$ across all slices are compared to the case when we adopt specific near optimum values of $\mu$ for each slice. It demonstrates that using a constant $\mu = 100$ can result in NRMSEs close to the optimal values.

Extending Joint MAPLE to whole-brain MR parameter mapping adds the slice dimension into the reconstruction and increases the computational burden. In the proposed framework, the volumetric MEMFA dataset (Fig. 2 part A) goes through coil-compression. In a uniformly random manner, a subset of slices of coil-compressed data is input to the pretrained joint ZS-SSL network (Fig. 2 part B). The network parameters are fine-tuned for one slice selected in each iteration. The validation loss is computed after the network is fine-tuned with all slices in the subset (Fig. 2 part C) to validate the generalizability of the network over all slices. In an inference step, all slices are reconstructed using the fine-tuned network. According to the framework shown in Fig. 1, the reconstructed volumes are input to the model block slice-by-slice to run the optimization process for mapping the MR parameters using a specific learning rate for each parameter. In the following, we detailed each step.

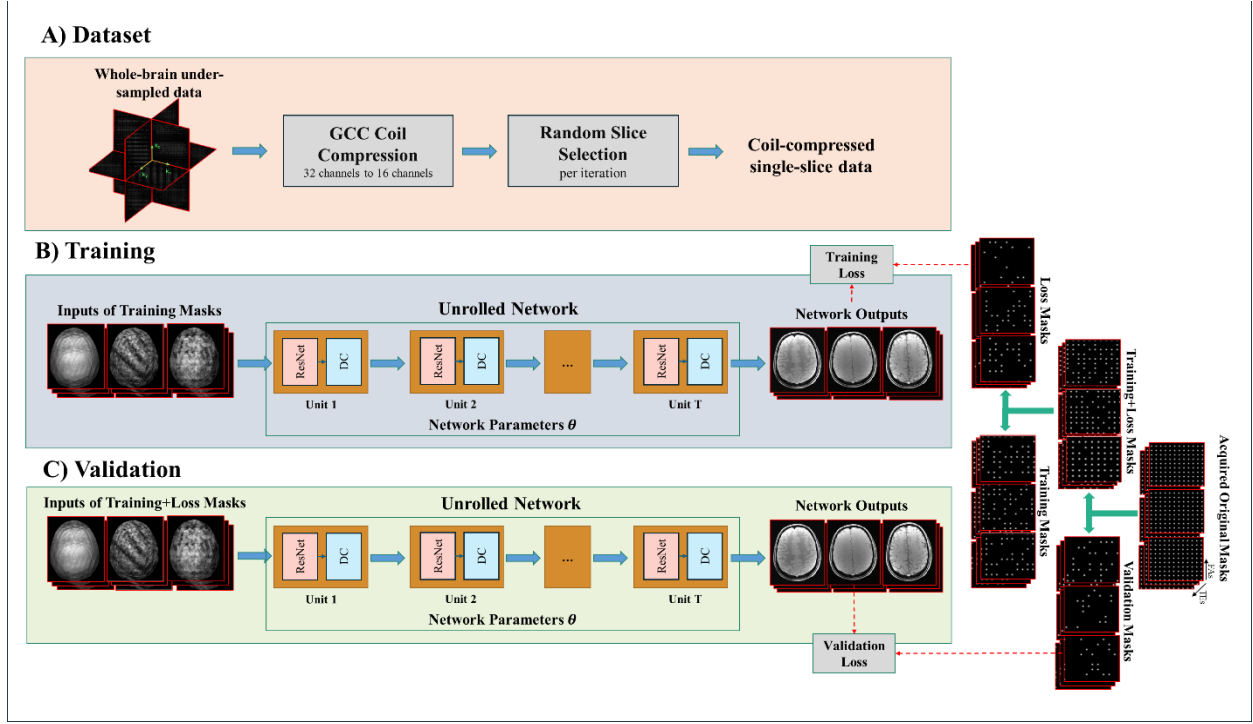

**Fig. 2. Joint ZS-SSL overall framework for whole-brain reconstruction.** Joint ZS-SSL is extended for whole-brain reconstruction of coil-compressed data with a random-slice selection manner. The joint ZS-SSL unrolls the ResNet-DC units with integrated convolution of different contrasts in the deep layers. It splits the acquired original k-space mask into training, loss and validation masks which are complementary across the contrasts. In each training iteration a random slice is selected to be fed into the network and the stopping criteria is extended to validate the generalizability of the network across different slices of a volume.

2.2. *Whole-brain data structure*

To enforce transfer learning, the proposed network needs MEMFA datasets from (at least) two different subjects. We assume that an initial subject has been scanned and the network is trained using this initial subject's data to yield the initial network parameters. Next, the network reconstructs the volume of the new subject's MEMFA dataset using the model trained from the prior subject as initialization.

### 2.3. Coil compression

We apply geometric-decomposition coil compression (GCC) (Zhang et al., 2013), which compresses the channels along a fully sampled readout dimension. Fig. 2 part A shows the coil compression as a prior step of injecting data into the network. This step is not necessarily required for the pre-training phase.

### 2.4. Random slice selection

The original ZS-SSL (Yaman et al., 2021) is designed to reconstruct a single-slice, single-contrast under-sampled k-space data. Joint MAPLE extended it to reconstruct different contrasts of the slice jointly with complementary under-sampling for each contrast (Bilgic et al., 2018). In the proposed FTL-Joint MAPLE we train joint ZS-SSL with multiple slices to reconstruct whole-brain data.

In practice, it is possible to make the architecture of joint ZS-SSL adapted to process a whole volume data. However, it comes with the cost of large memory usage and time-consuming iterations. On the other hand, a ZS-SSL network trained by a single slice can infer other slices from their under-sampled k-space data. This feature of ZS-SSL can offer a significant reduction in reconstruction time. But such a network that is trained by a single slice may lack sufficient generalizability over the entire volume. To address these, we use random slice selection to avoid exhaustive training of the network with all slices in each iteration. Additionally, the network encounters all slices during the training to keep the generalizability over the whole volume.

ZS-SSL terminates the training when the validation loss starts to increase. It assumes that after a certain number of training steps the trained network could not offer better generalizability in terms of validation loss for new k-space locations. But in a multi-slice setting, the concept of generalizability shifts from the performance of the network over unseen locations of k-space for a single slice to unseen slices of the volume. Therefore, the stopping criteria should be changed based on this new concept. As such, we average the validation loss values calculated by the network after a certain number of iterations and then evaluate the stopping criteria. This way, termination criterion for the training (or fine-tuning) iterations is evaluated for a set of slices instead of a single slice.

### 2.5. Transfer learning reconstruction with joint ZS-SSL

The proposed framework employs a simple transfer learning approach to extend the utility of a pre-trained joint ZS-SSL for all new datasets. FTL-Joint MAPLE exploits transfer learning to accelerate the convergence of the reconstruction iterations by initialization of the network parameters using data acquired on a prior subject. Since the training epochs in joint ZS-SSL are time consuming, reduction in the number of epochs results in faster training of the network for the new subject.

*2.6. Parameter-specific learning rates*

The $T_1$, $T_2^*$, proton density and frequency maps estimated in FTL-Joint MAPLE have different dynamic ranges. For example, the values of $T_1$ map are ~20 times greater than the values of $T_2^*$ on average. Proton density is complex-valued while other maps are real-valued. In the proposed FTL-Joint MAPLE, the learning rates are specific to $T_1$, $T_2^*$ proton density and frequency map, and held constant throughout the iterations.

*2.7. Dataset*

Fully sampled in-vivo 3D whole-brain datasets from two different subjects were acquired for pretraining and fine-tuning steps. A Siemens 3T Prisma scanner with a 32-channel head-coil is used. Their field-of-view (FOV) are 224 mm × 192 mm × 160 mm and 224 mm × 192 mm × 144 mm. The matrix sizes are 224×192×80 and 224×192×72. The repetition time (TR) is 34 ms for one subject and 33 ms for the other subject. The slice thickness is 2 mm. Each dataset consists of six different echo times (TEs) for each of the three flip angles (FAs) of 4, 10, 16 degrees. The first TE is 3.6 ms and the difference between each is ΔTE = 5 ms. Scan time is 6:50 min and 7:38 min for fully sampled data per flip angle for subject 1 and 2 respectively. An example slice of two datasets are shown in supplementary Fig. S1 and Fig. S2 with their detailed specifications.

3. **Results**

The fully sampled whole-brain data was retrospectively under-sampled using a 2D uniform sampling mask. The sampling scheme uses a specific mask for each contrast in a complementary manner to cover more locations of the k-space data. In all figures we show the overall pattern of the under-sampling mask on the top right of the figure. The column bar beside each mask shows a part of the collective mask generated by adding three different masks together with different TEs/FAs to show the masks' collective frequency coverage. The acceleration rate is R = 12 (4x3). To evaluate the accuracy of estimated parameter maps, normalized RMSEs are measured between the generated parameter maps from the fully sampled data and the estimated maps from under-sampled data, after applying a brain mask.

To control the effect of very large $T_1$ and $T_2^*$ values (due to extremely small $R_1$ and $R_2^*$ values) in the metrics, we saturated their values within ranges of 0-2500 ms and 0-100 ms for the purpose of NRMSE computation, respectively. To evaluate the performance of the methods in terms of the reconstruction time, we use whole-brain reconstruction time (WBRT) metric which combines different processing times including initializations, multi-contrast reconstruction and parameter estimation for all slices.

Codes are implemented on a combination of CPU (Core i7 intel with 16G system RAM using a personal computer) and GPU (A100 with 40G GPU RAM using Google Collaboratory) processors.

In the first experiment (Fig. 3), we compared the mapping performance of the proposed FTL-Joint MAPLE with SENSE (Pruessmann et al., 1999), a structured low rank technique LORAKS (Haldar, 2013; Kim et al., 2017) followed by least-squares parameter mapping and the original Joint MAPLE (Heydari et al., 2024) in terms of accuracy and reconstruction time. SENSE and LORAKS use individual signal models (Eqs. (3) and (4)) and they do not estimate $T_1$ and $T_2^*$ jointly. Therefore, given three flip angles and six echo times in the acquired MEMFA dataset, these methods estimate three separate $T_2^*$ and six separate $T_1$ maps for each slice. In Fig. 3, we showcased the results for a single slice (number 25) of the whole-brain dataset. For SENSE and LORAKS, shown results are the average maps and the metrics are the average values across different TEs/FAs of MEMFA dataset. Joint MAPLE and FTL-Joint MAPLE reconstruct a single estimation of $T_1$, $T_2^*$, proton density (P) and frequency map ($\Delta B$) for a given MEMFA dataset which are shown in the figure. The k-space data in FTL-Joint MAPLE is compressed to 16 channels (as part of the proposed framework) while other techniques use a 32-channel k-space data. The same architecture of ZS-SSL network is used for both Joint MAPLE and FTL-Joint MAPLE.

SENSE and LORAKS follow a two-step manner for MR parameter mapping: 1) reconstruction of under-sampled contrasts and 2) least square parameter fitting using signal models. For SENSE and LORAKS, we used CPU for the reconstruction step and GPU for the parameter fitting. Joint MAPLE and FTL-Joint MAPLE were fully implemented on GPU for faster computation. The WBRT for the FTL-Joint MAPLE was obtained as the average of 10 runs with the standard deviation of 3.15 minutes.

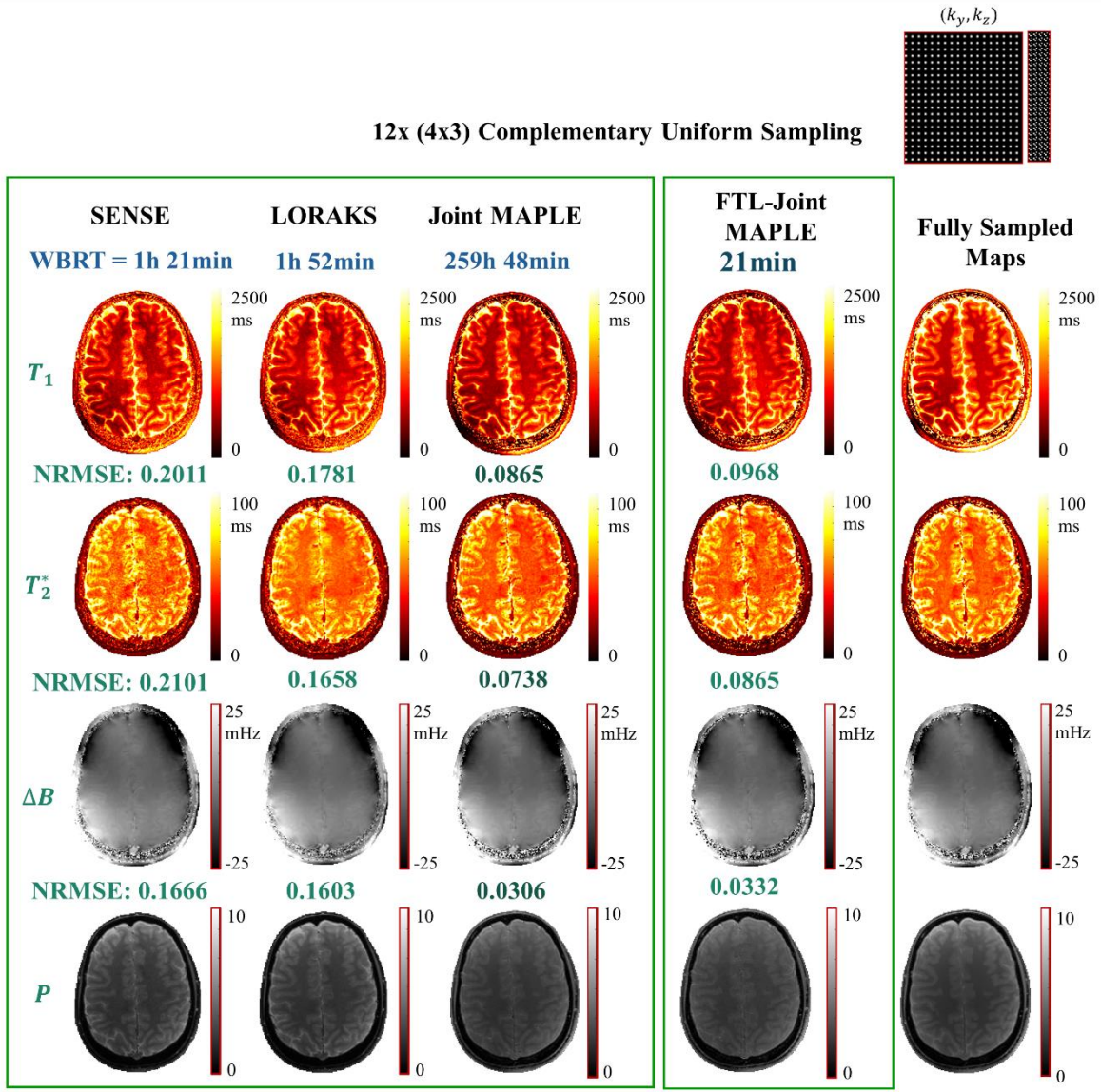

**Fig. 3. FTL-Joint MAPLE vs parallel imaging techniques.** The performance of the proposed FTL-Joint MAPLE vs SENSE, LORAKS and Joint MAPLE in terms of whole-brain reconstruction time (WBRT) and the showcased single slice normalized RMSE. The acceleration rate is R = 12 (4x3) with a complementary uniform sampling scheme. The maps and the metrics for SENSE and LORAKS are the average across TEs/FAs. Reported WBRT for FTL-Joint MAPLE is the average of 10 runs. The proposed FTL-Joint MAPLE is significantly faster than the original Joint MAPLE with close NRMSEs in all parameters. It also outperforms SENSE and LORAKS significantly in both accuracy and reconstruction time for all parameters.

As shown in Fig. 3, the proposed FTL-Joint MAPLE can estimate $T_1$, $T_2^*$, proton density and frequency maps of a whole-brain in 21 minutes on average. This is a large reduction in comparison to the original Joint MAPLE which needs approximately 10 days for the same task, provided that all slices are processed sequentially. The measured accuracy (NRMSEs) demonstrates that this reduction in reconstruction time does not come at the cost of the fidelity of the maps. FTL-Joint MAPLE is faster than SENSE and LORAKS with better mapping performance in all parameters, but we note that SENSE/LORAKS use CPU processing for their image reconstruction step. The supplementary Fig. S3 shows the results of this experiment for a different subject which confirms the results summarized in Fig. 3.

The proposed FTL-Joint MAPLE uses a random slice selection scheme to train the network parameters with different slices of the volume. However, joint ZS-SSL trained by a single slice can also reconstruct other unseen slices of the same subject, but this may come at the cost of lack of generalizability. Fig. 4 shows the results of the next experiment where we compare the generalizability performance of the original Joint MAPLE and the proposed FTL-Joint MAPLE in parameter mapping of different slices. The joint ZS-SSL network in Joint MAPLE is trained by a single slice (number 15) and using the trained network all other slices are inferred to reconstruct a whole-brain volume. FTL-Joint MAPLE trains the network using all slices. Joint MAPLE and FTL-Joint MAPLE show a close performance for slice 15. But for the other two slices, FTL-Joint MAPLE outperforms Joint MAPLE in mapping of most parameters, indicating improved generalizability.

In the next experiment, the $T_1$ and $T_2^*$ mapping performance of the proposed FTL-Joint MAPLE is compared to the standard and state-of-the-art parallel imaging techniques of SENSE, joint ZS-SSL and LORAKS for different slices. Joint ZS-SSL and LORAKS jointly reconstruct the multi-contrast data across different TEs /FAs. As shown in Fig. 5, the proposed framework outperforms the other methods in all parameters across different slices. In Fig. 5, the average maps are shown for parallel imaging techniques. Joint ZS-SSL is taken as a state-of-the-art parallel imaging technique followed by least squares parameter fitting to estimate the parameters separately. The 16-channel k-space data is used for all techniques. Lower slices are more challenging for all reconstruction techniques in comparison to upper slices. Supplementary to this experiment, supplementary Fig. S4 demonstrates the results for a different subject reporting the ability of FTL-Joint MAPLE to outperform other techniques for different slices.

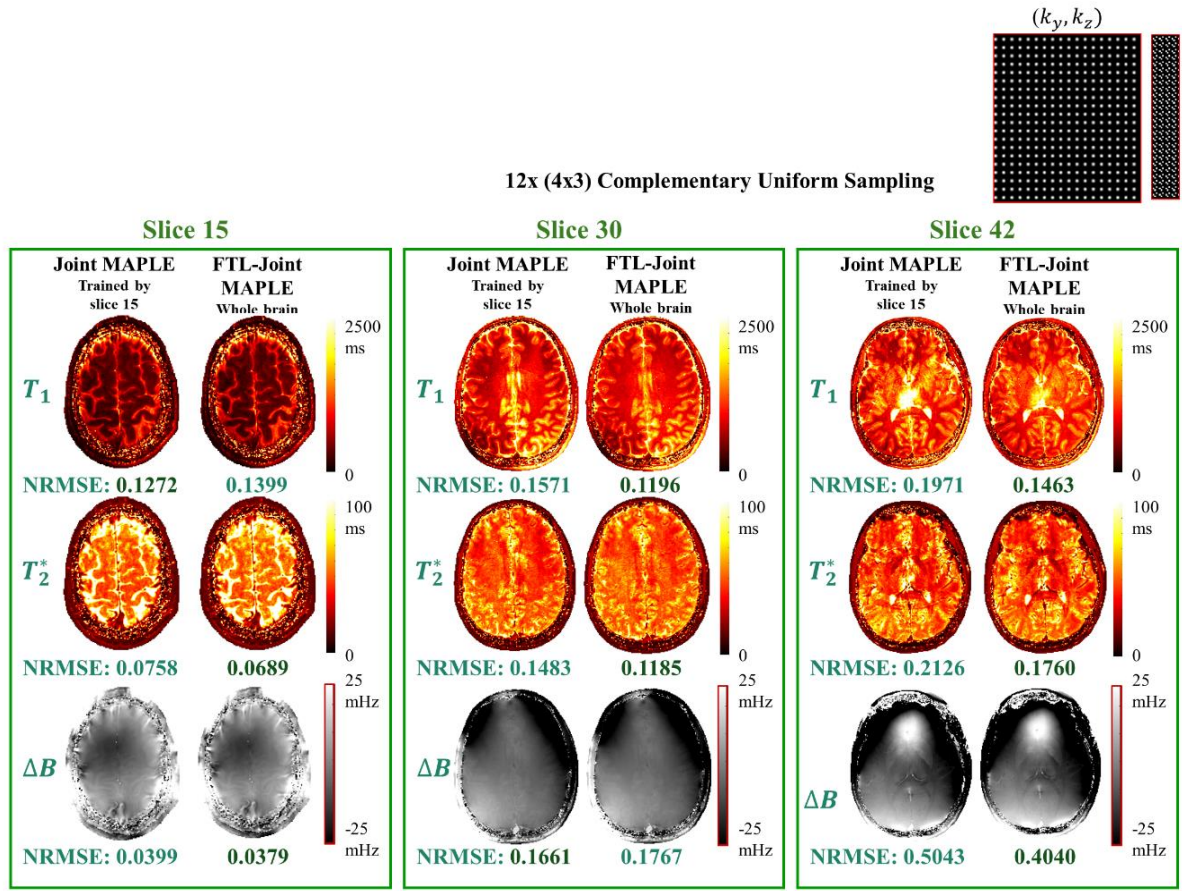

**Fig. 4. Generalizability evaluation.** Generalizability comparison of Joint MAPLE trained by slice 15 and FTL-Joint MAPLE trained by the random slice selection manner. The acceleration rate is R = 12 (4x3) and a complementary uniform sampling is used. For slice 15, two methods perform closely. For slices 30 and 42 which are unseen for Joint MAPLE, the mapping performance of the proposed FTL-Joint MAPLE is better than Joint MAPLE in all parameters which is an indicator of better obtained generalizability with whole-brain training.

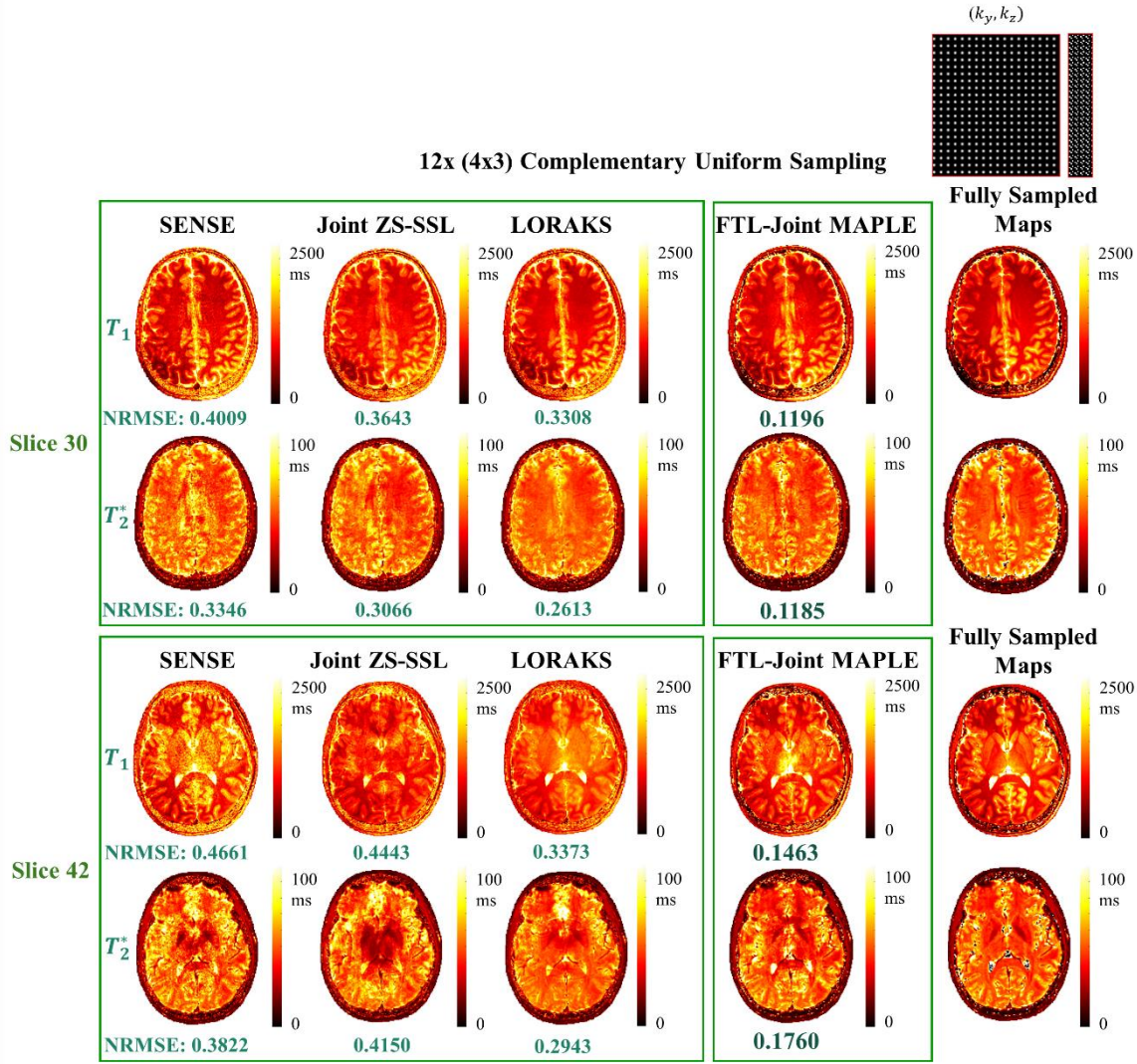

**Fig. 5. Mapping performance of FTL-Joint MAPLE vs SENSE, joint ZS-SSL and LORAKS for different slices.** The acceleration rate is R = 12(4x3) with a complementary uniform sampling scheme. The results for SENSE, joint ZS-SSL and LORAKS are the average results across TEs/FAs. FTL-Joint MAPLE outperforms other techniques for different slices in all parameters and it retains its mapping performance while accelerating the reconstruction process. The reconstruction of lower slices is more challenging for all techniques.

Axial, sagittal and coronal views of the reconstructed whole-brain $T_1$ and $T_2^*$ is shown in Fig. 6. The reported NRMSE for each map is the whole-brain NRMSE after applying a brain mask. The increased $B_0$ frequency inhomogeneity for lower slices adversely affects the reconstruction process and this effect is clearer in sagittal and coronal views. The $T_1$ and $T_2^*$ maps of the middle slices of each view are shown in the figure. The supplementary Fig. S7 demonstrates an axial, sagittal and coronal view of the reconstructed maps for SENSE and

LORAKS with $T_1$ NRMSEs of 0.3011 and 0.2389 and $T_2^*$ NRMSEs of 0.4063 and 0.3502 for SENSE and LORAKS respectively.

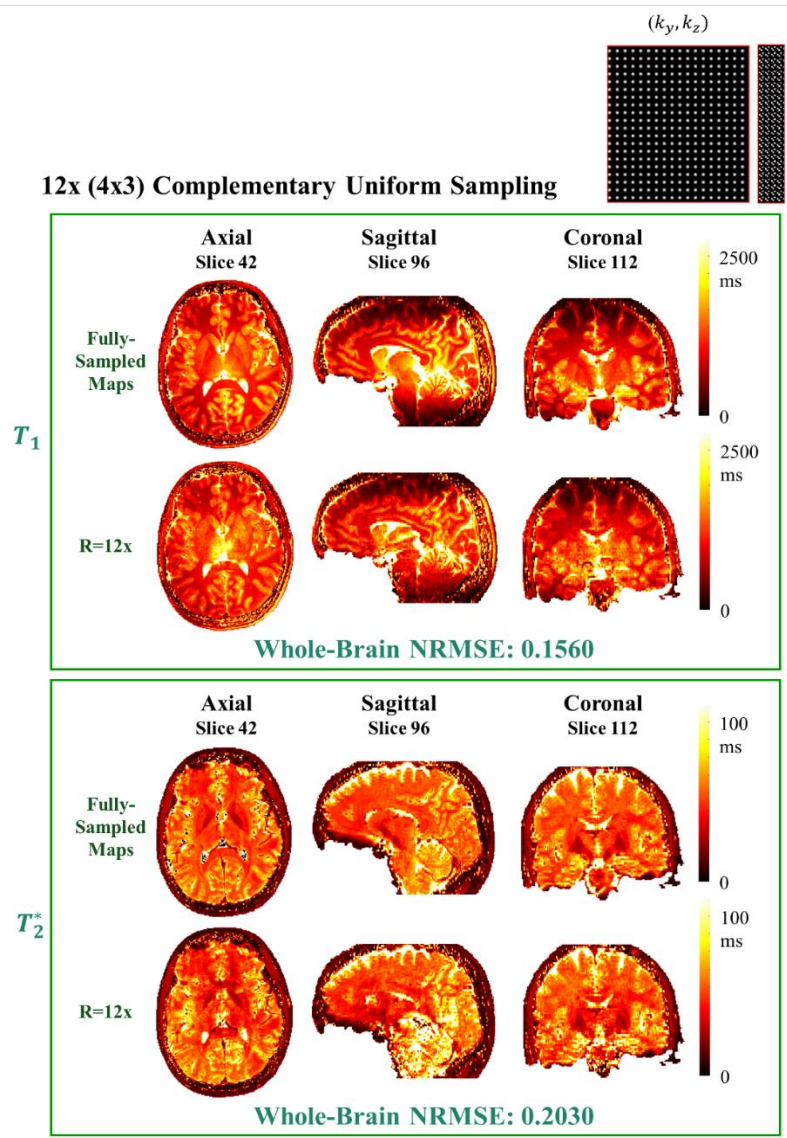

**Fig. 6. A three-view (axial, sagittal and coronal) of the whole-brain reconstructed $T_1$ and $T_2^*$ maps which showcases the middle slice of each view.** The acceleration rate is R = 12 (4x3) with a complementary uniform pattern. The reported NRMSEs are the whole-brain NRMSEs after applying a brain mask.

3.1. *Ablation Study*

FTL-Joint MAPLE incorporates whole-brain reconstruction, GCC, parameter-specific learning rates and transfer learning approach to create a fast version of Joint MAPLE. In an ablation study, we evaluate the impact of each contribution on NRMSE and WBRT metrics in a step-by-step manner. Fig. 7 summarizes the results of each step for example slice of 25.

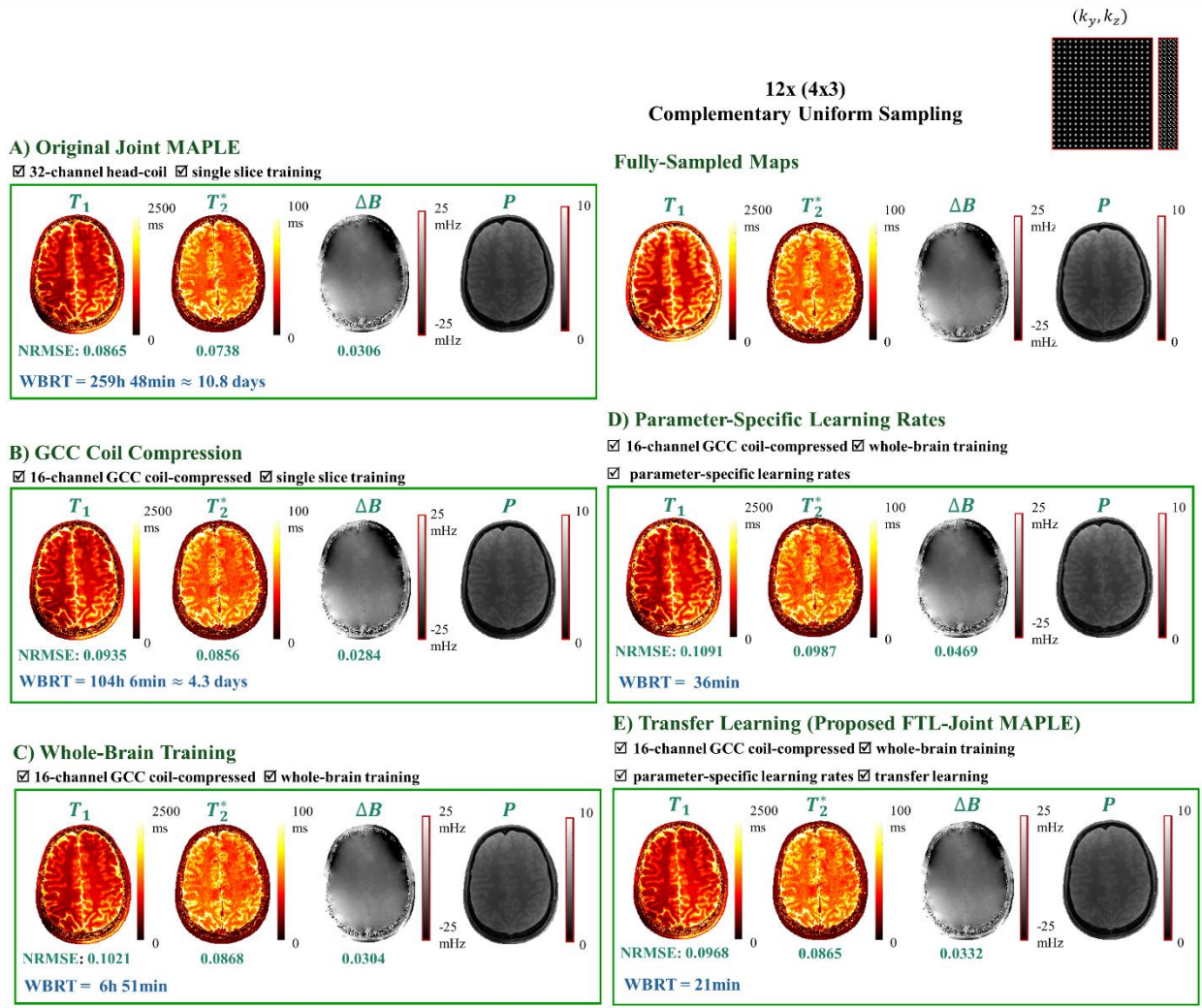

**Fig. 7. Ablation study of the effect of each proposed contribution on the mapping metrics in successive parts.** The acceleration rate is R = 12 (4x3) with a complementary uniform pattern. A) The original Joint MAPLE with no acceleration contribution which shows the best overall mapping performance in terms of the measured errors. B) Joint MAPLE performance on the coil-compressed version of the same dataset. The significant reduction in reconstruction time shows the impact of the size of dataset in the run time of the framework. C) Incorporating whole-brain reconstruction and coil compression into Joint MAPLE reduces the reconstruction time from a few days to ~ 7 hours. D) Optimization of whole-brain Joint MAPLE with parameter-specific learning rates using coil-compressed dataset which decreases the reconstruction time to less than one hour. E) The proposed FTL-Joint MAPLE with added transfer learning with the best reconstruction time and mapping performance comparable with the original Joint MAPLE (part A).

The original Joint MAPLE (Fig. 7 part A) has the best overall mapping performance in comparison to other versions which comes at the cost of the longest reconstruction time. GCC (Fig. 7 part B) results in a significant reduction in reconstruction time (2x faster). Adding the whole-brain reconstruction to the framework (Fig. 7 part C) reduces the WBRT from a few days to approximately 7 hours. However, this comes with the cost of a slight increase in NRMSEs. Incorporating the parameter-specific approach (Fig. 7 part D) reduces WBRT to almost half an hour. Finally, employing transfer learning (Fig. 7 part E) offers the final reduction in WBRT to 21 minutes. In addition to reconstruction time, using transfer learning yields improvements in mapping performance of the framework as NRMSE values show better values for all parameters than the last two previous versions in C and D.

The role of the fine-tuning phase in the proposed framework is also proved to be critical for reaching high fidelity maps of the new subject. The supplementary Fig. S5 indicates that removing the fine-tuning phase can help the speed of the reconstruction. However, the estimated parameters can be degraded significantly in terms of the NRMSE metric.

4. **Discussion**

The previously proposed Joint MAPLE exploits the rich information in MEMFA data and combines parallel imaging, model-based and deep learning approaches. It yields high fidelity MR parameter estimation at high accelerations. However, it is not generalizable to volumetric reconstructions due to its high computational complexity. FTL-Joint MAPLE incorporates four strategies into the Joint MAPLE framework that yield significant gains in reconstruction time of a whole-brain MEMFA dataset.

The FTL-Joint MAPLE framework consists of two main blocks, recon and model blocks. It adopts a random slice selection strategy to avoid memory limitations and prohibitive computation costs. The larger proportion of the reconstruction time (~70%) is due to the joint ZS-SSL reconstruction in the recon block. So, faster reconstruction speeds can be achieved by new strategies that make ZS-SSL more efficient. For example, increasing the generalization capability of the joint ZS-SSL for new slices or new subjects can be helpful for removing the fine-tuning phase in the FTL-Joint MAPLE framework. The results in the supplementary Fig. S5 indicate that removing the fine-tuning phase from the current framework of FTL-Joint MAPLE accelerates the reconstruction but this comes at the cost of accuracy of parameter mapping.

MR parameter mapping with parallel imaging techniques includes an image reconstruction phase followed by a parameter fitting step like dictionary matching or least square optimization. LORAKS is more time consuming than SENSE in the image reconstruction phase. However, the calculated WBRT for them in Fig. 3 is dominated by the parameter fitting step which is more time consuming than the image reconstruction phase. In the experiments, SENSE and LORAKS take ~6 and ~38 minutes on average for whole-brain image reconstruction across different TEs/FAs respectively, while their parameter fitting step takes ~76 and ~74 minutes for SENSE and LORAKS respectively for $T_1$, $T_2^*$, proton density and frequency map estimation.

Adjustment of the regularization terms of $\mu$ and $\mu_R$ in Eq. (8) is effective in achieving the best mapping performance. The heuristic method (Heydari et al., 2024) and automatic parameter tuning techniques (Iyer et al., 2020; Toma and Weller, 2020; Weller et al., 2014) can be applied based on the minimization of the NRMSE metric for a single slice. However, a main challenge is the fact that the optimum regularization weights could be different for different slices and subjects. An empirical rule that can be used is that for slices with better field homogeneity, smaller values of $\mu$ can be selected to increase the effect of the quality reconstruction of the contrasts. The reported results are using near optimum $\mu$ values for each slice. However, a constant value of $\mu = 50$ or 100 for all slices result in NRMSEs close to the optimum values (supplementary Fig. S8).

The dataset used in the experiments includes 80 slices at 2 mm thickness. However, the reconstruction time does not change linearly as a function of the number of slices. As mentioned, the recon block is responsible for almost 70% of the reconstruction time which uses a random slice selection manner. A random subset with a certain number of slices (e.g. 10 slices) which is independent of the number of all slices is input to each epoch. The validation criterion is evaluated in each epoch. Therefore, the stopping condition could be met in each epoch independent of the number of slices in the volume. The supplementary Fig. S3 shows the results for another volume with 72 slices. The calculated WBRT is close to the volume with 80 slices which has more slices to process.

The reconstruction time of FTL-Joint MAPLE is also not linearly related to the number of TEs or FAs. Separate mapping of $T_1$ and $T_2^*$ with the FTL-Joint MAPLE framework using Eq. (3) and Eq. (4) needs three and six contrasts respectively to process. Consequently, there will be a reduction in whole-brain reconstruction time due to less data to process. However, this reduction in reconstruction time is not linear with the number of TEs/FAs in the dataset. Supplementary Fig. S6 shows the WBRTs and NRMSEs for separate reconstruction of $T_1$ and $T_2^*$ with the FTL-Joint MAPLE framework.

Transferring the large MEMFA data to the system memory is a time-consuming part dependent on the system specifics and is not included in the reported metrics for any of the techniques and experiments. The acquired dataset requires the calibration operations like coil sensitivity mapping and coil compression before the reconstruction steps. The time required for these steps were not included in the reported WBRT metrics either.

We note that FTL-Joint MAPLE also suffers from model mismatch similar to the original framework. The images synthesized from the estimated parameter maps do not match the acquired MEMFA data perfectly, presumably due to model mismatch in the complex multi-parametric signal model. This problem causes a baseline error in synthesized MEMFA images as the output of the model block, but it does not affect the overall performance of the proposed framework. Fig. 8 shows that the synthetic images generated by FTL-Joint MAPLE are affected by the model mismatch presumably due to the complexity of the multi-parametric signal model of Eq. (2). Therefore, they cannot compete with the synthesized images from parallel imaging techniques in terms of NRMSE, while FTL-Joint MAPLE shows a significantly better parameter mapping performance.

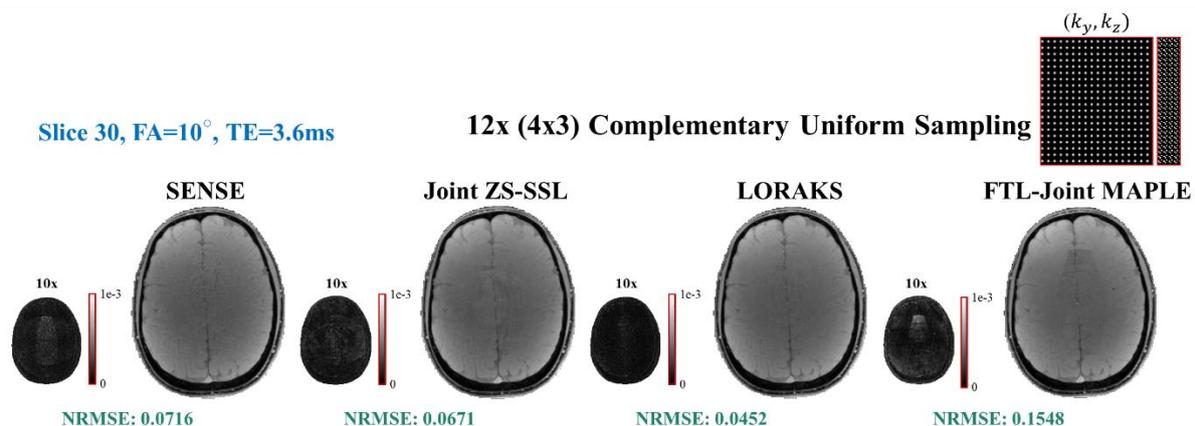

**Fig. 8. Synthetic images generated by parallel imaging techniques vs FTL-Joint MAPLE.** Eq. (3) is used for SENSE, joint ZS-SSL and LORAKS, and FTL-Joint MAPLE uses Eq. (2) as the signal model. The generated contrast related to FA = $10^o$ and TE = 3.6ms for slice 30 is showcased with 10x scaled error image for each technique. NRMSEs are the average values across different TEs/FAs of slice 30. While its outperformance in MR parameter mapping, the synthetic images generated by FTL-Joint MAPLE show higher error values due to the model mismatch effect.

The visual artifacts in the showcased synthetic image of FTL-Joint MAPLE, in Fig. 8, are not propagated to the estimated parameter maps (please see Fig. 5, estimated maps of FTL-

Joint MAPLE for slice 30). This is mostly because the proposed framework utilises all information in all contrasts. However, for parallel imaging techniques, the averaging operation during parameter mapping softened the artifacts' propagation from their reconstruction step (Fig. 5).

### 4.1. *Limitations*

Like the original Joint MAPLE, the proposed framework is applicable to qMRI pulse sequences with explicit, closed-form signal models. However, popular and efficient sequences such as MR Fingerprinting (Ma et al., 2013), 3D-QALAS (Kvernby et al., 2014), 3D-EPTI (Wang et al., 2022) and MR Multitasking (Christodoulou et al., 2018) often do not possess an explicit signal model. To address this problem, one can replace the closed-form signals with a subspace approach where a low-rank approximation to the Bloch equation-based simulation of the specific qMRI sequence's dynamics can be used (Cao et al., 2022; Tamir et al., 2017). Instead of the parameter maps, subspace coefficients can be estimated within the Joint MAPLE framework.

The performance of the proposed framework can be affected by additional factors in the data acquisition stage. Increased inhomogeneities in $B_0$ for lower slices with air/tissue interfaces impose a larger dynamic range for frequency maps which may be a reason for challenging reconstruction of the lower slices. Additionally, the multi-flip angle acquisition makes data prone to potential patient motions across flip angles. In a prospectively under-sampled acquisition, the achievable high acceleration rates by FTL-Joint MAPLE can mitigate the sensitivity to patient motion. Moreover, techniques like rigid motion correction across contrasts can be incorporated into the framework to enhance the motion robustness.

The resolution of the dataset used in this study is limited to 1mm x 1mm x 2mm with 6:50 minutes acquisition time for a single flip angle to minimize the risk of involuntary subject motion. However, isotropic resolution for many applications like QSM is desirable (QSM Consensus Organization Committee, 2024) which requires a scan time of almost 45 minutes for three flip angles for fully-sampled data. A practical alternative could be to collect mildly accelerated data (e.g. R = 2x) at 1mm iso. followed by e.g. GRAPPA (Griswold et al., 2002) reconstruction to estimate "fully-sampled" data for retrospective accelerations.

## 5. Conclusions

In this work we proposed a fast extension of Joint MAPLE parameter mapping framework which is a synergistic combination of parallel imaging, model-based and deep learning approaches for multi-parametric mapping of $T_1$, $T_2^*$, proton density and frequency maps at high acceleration rates with high fidelity.

The proposed FTL-Joint MAPLE accelerates the reconstruction process of Joint MAPLE while retaining its high mapping performance. It addresses the increased size of the whole-brain dataset, complexity of optimization terms and multi-parametric nature of the framework with four effective strategies. The proposed FTL-Joint MAPLE incorporates coil compression, whole-brain reconstruction, parameter-specific learning rates and a transfer learning approach to the original framework, thus yielding significant reduction in processing time. Exploiting these contributions enables FTL-Joint MAPLE to be up to 700 times faster than the original Joint MAPLE when processing a whole-brain MEMFA dataset. It reduces the parameter mapping time of volumetric data to 21 minutes on average which makes it more applicable in clinical routines and research studies.

It was shown that FTL-Joint MAPLE has a better generalizability across different slices in comparison to Joint MAPLE. The fidelity of estimated maps is close to the results of Joint MAPLE in terms of NRMSE and approximately 2-fold better than other techniques on the average.

## Acknowledgements

This work was supported by research grants NIH R01 EB028797, P41 EB030006, U01 EB026996, R03 EB031175, R01 EB032378, UG3 EB034875, R21 AG082377, R01 EB034757, S10 OD036263;

NVidia Corporation for computing support and National Research Foundation of Korea (NRF);

Grant funded by the Korea government (MSIT) (No. 2022R1F1A1074786).